\documentclass[twocolumn]{aastex61}

\pdfoutput=1 
\usepackage{amsmath,amstext}
\usepackage[T1]{fontenc}
\usepackage[figure,figure*]{hypcap}
\usepackage{natbib}
\usepackage{xspace}

\usepackage{hyperref}
\usepackage{cleveref}

\newcommand{\chandra}{{\it Chandra}\xspace}
\newcommand{\hst}{{\it HST}\xspace}

\def\ergs{\hbox{${\rm erg\, s^{-1}}$}\xspace}
\def\kms{\hbox{${\rm km\, s^{-1}}$}\xspace}

\def\Mgb{\hbox{${\rm Mg}\, b$}\xspace}
\def\nx{\hbox{$n_{x}$}\xspace}

\def\Msun{\hbox{$\thinspace M_{\odot}$}\xspace}

\def\LKsun{\hbox{$\thinspace L_{K\odot}$}\xspace}

\def\lx{\hbox{$\thinspace L_{x}$}\xspace}

\received{February 23, 2017}
\accepted{May 01, 2017}
\submitjournal{ApJ}

\shorttitle{LMXB constraints on IMF variations}
\shortauthors{Peacock et al.}

\begin{document}

\title{Further constraints on variations in the IMF from LMXB populations}

\correspondingauthor{Mark B. Peacock}
\email{mpeacock@msu.edu}

\author{Mark B. Peacock}
\affiliation{Department of Physics and Astronomy, Michigan State University, East Lansing, MI 48824, USA}

\author{Stephen E. Zepf}
\affiliation{Department of Physics and Astronomy, Michigan State University, East Lansing, MI 48824, USA}

\author{Arunav Kundu}
\affiliation{Eureka Scientific, Inc., 2452 Delmer Street, Suite 100 Oakland, CA 94602, USA}

\author{Thomas J. Maccarone}
\affiliation{Texas Tech University, Lubbock, TX 79409, USA}

\author{Bret D. Lehmer}
\affiliation{University of Arkansas, 226 Physics Building, 835 West Dickson Street, Fayetteville, AR 72701, USA}

\author{Claudia Maraston}
\affiliation{Institute of Cosmology and Gravitation, Dennis Sciama Building, Burnaby Road, Portsmouth PO1 3FX, UK}

\author{Anthony H. Gonzalez}
\affiliation{University of Florida, Gainesville, FL 32611, USA}

\author{Rafael T. Eufrasio}
\affiliation{University of Arkansas, 226 Physics Building, 835 West Dickson Street, Fayetteville, AR 72701, USA}

\author{David A. Coulter}
\affiliation{University of California Santa Cruz, Santa Cruz, CA 95064, USA}

\begin{abstract}
\label{sec:abstract}

We present constraints on variations in the initial mass function (IMF) of nine local early-type galaxies based on their low mass X-ray binary (LMXB) populations. Comprised of accreting black holes and neutron stars, these LMXBs can be used to constrain the important high mass end of the IMF. We consider the LMXB populations beyond the cores of the galaxies ($>0.2R_{e}$; covering $75-90\%$ of their stellar light) and find no evidence for systematic variations of the IMF with velocity dispersion ($\sigma$). We reject IMFs which become increasingly bottom heavy with $\sigma$, up to steep power-laws (exponent, $\alpha>2.8$) in massive galaxies ($\sigma>300$~\kms), for galactocentric radii $>1/4\ R_{e}$. Previously proposed IMFs that become increasingly bottom heavy with $\sigma$ are consistent with these data if only the number of low mass stars $(<0.5\Msun$) varies. We note that our results are consistent with some recent work which proposes that extreme IMFs are only present in the central regions of these galaxies. We also consider IMFs that become increasingly top-heavy with $\sigma$, resulting in significantly more LMXBs. Such a model is consistent with these observations, but additional data are required to significantly distinguish between this and an invariant IMF. For six of these galaxies, we directly compare with published ``IMF mismatch" parameters from the Atlas3D survey, $\alpha_{dyn}$. We find good agreement with the LMXB population if galaxies with higher $\alpha_{dyn}$ have more top-heavy IMFs -- although we caution that our sample is quite small. Future LMXB observations can provide further insights into the origin of $\alpha_{dyn}$ variations. 

\end{abstract}


\section{Introduction}
\label{sec:intro}

The stellar initial mass function (IMF) has an important impact on a wide range of astrophysics, from understanding the collapse of molecular clouds to stellar feedback and large scale structure formation. Unfortunately, constraining the form of the IMF is observationally very challenging, particularly beyond the resolved stellar populations of the Local Group. For this reason a universal IMF, based on that observed in the Milky Way, is commonly adopted for all galaxies. This has a form similar to that presented by \citet{Kroupa01} and \citet{Chabrier03} and can be described by the broken power-law $dN/dm \propto m^{-\alpha}$, where $\alpha_{2}=2.3$ for stars with $m>0.5\Msun$ and $\alpha_{1}=1.3$ for lower mass stars with $0.5>m>0.08\Msun$. There is no evidence for strong variation in the Milky Way's IMF \citep[e.g.][]{Bastian10}. However, observations of the unresolved stellar populations of local early-type galaxies suggests that their IMFs may not be similar, but rather vary systematically with galaxy mass. 

The IMF of unresolved stellar populations can be inferred from modeling the integrated emission. Of particular importance are IMF-sensitive spectral features, which can probe the ratio of low- to high-mass stars \citep{Cohen78,Faber80,Carter86,Couture93}. Such work has shown that, as the mass of a galaxy increases, the strength of the giant sensitive Ca~{\sc ii} triplet decreases \citep{Saglia02, Cenarro03} while the strength of dwarf sensitive features such as Na~{\sc i} and the Wing-Ford molecular FeH band increases \citep{vanDokkum10, vanDokkum11, Smith12, Spiniello12, Ferreras13, Spiniello14, Smith15, LaBarbera16}. This suggests that the IMF may become increasingly bottom heavy with increasing galaxy mass, although correlations with other parameters, such as metallicity or $\alpha$-abundance, have also been proposed \citep[e.g.][]{Conroy12b, McDermid14, LaBarbera15, Martin_Navarro15b, vanDokkum16}. Similar conclusions have been drawn from fitting stellar population synthesis models to the full spectra of galaxies \citep{Conroy12b, Ferreras13, LaBarbera13}. 

\citet{Spiniello15} demonstrated that the inferred variability of the IMF is robust to the choice of stellar populations models. However, inferring the IMF from such observations is complex and \citet{Spiniello15} do demonstrate significant differences in the inferred magnitude of the IMF variations. Alternative explanations for some of these spectral features have also been suggested -- such as including binary stars \citep[although this cannot explain Ca variations,][]{Maccarone14}. Recent work has also suggested significant radial variability in the IMF, with the central regions of the most massive galaxies having the most extreme IMFs \citep[e.g.][]{Martin_Navarro15a, McConnell16, vanDokkum16, LaBarbera16, LaBarbera17}. 

Observations of dwarf galaxies further support the idea of an IMF that becomes increasingly bottom heavy with increasing galaxy mass. Direct stellar counts of the Milky Way satellites Hercules and Leo IV have shown that they have flatter than Kroupa IMFs, consistent with their low masses \citep{Geha13}. Additionally, ultra-compact dwarf galaxies have also been shown to have high $M/L$ ratios that could also be explained by them having a relatively flat IMF \citep[although dark matter fractions could also explain these data; e.g.][]{Hasegan05, Mieske08a, Mieske08b}. 

An independent method of inferring the IMF of a galaxy is to determine the mass to light ($M/L$) ratio via either gravitational lensing or dynamical measurements and compare it to that based on stellar population synthesis models. \citet{Treu10} studied 56 strong gravitational lensing galaxies and concluded that either the IMF or the dark matter halo must vary. Assuming similar dark matter profiles, these observations suggest that the IMF of the most massive galaxies must have significantly higher $M/L$ than that of the Milky Way \citep{Treu10, Auger10}. However, it should also be noted that observations of some massive ($\sigma>300$~\kms), strong lensing galaxies were found to be consistent with having a Milky Way like IMF and inconsistent with Salpeter or steeper IMFs \citep{Smith13, Smith15, Newman16}.

\begin{deluxetable*}{clccccccccccccc}[t!]
\tablecaption{Galaxy sample \label{tab:gal_data}}
\tablecolumns{15}
\tablenum{1}
\tablewidth{0pt}
\tablehead{
\colhead{Name} & \colhead{Type\tablenotemark{a}} & 
\colhead{D\tablenotemark{b}} & \colhead{ref\tablenotemark{b}} & 
\colhead{$\sigma$\tablenotemark{c}} & \colhead{ref\tablenotemark{c}} & 
\colhead{\Mgb\tablenotemark{d}} & \colhead{$[Z/H]$\tablenotemark{d}} & \colhead{$[\alpha/Fe]$\tablenotemark{d}} & 
\colhead{$R_{e}$\tablenotemark{e}} & 
\colhead{$r_{in}$\tablenotemark{f}} & \colhead{$r_{ext}$\tablenotemark{g}} & \colhead{$e$\tablenotemark{g}} & 
\colhead{$L_{K}$\tablenotemark{g}} & \colhead{$f_{K}$\tablenotemark{g}} 
\\
\colhead{(NGC)} & \colhead{} & 
\colhead{(Mpc)} & \colhead{} & 
\colhead{(\kms)} & \colhead{} & 
\colhead{} & \colhead{} & \colhead{} & 
\colhead{($\arcsec$)} & 
\colhead{($\arcsec$)} & \colhead{($\arcsec$)} & \colhead{} & 
\colhead{($\times10^{10}\LKsun$)} & \colhead{} 
}
\startdata
  1399 & E1   & 20.0 & 1 & 280 & 2 &    --    &    --  &    --   & 48.6 & 10    & 220.2 & 0.00 & 25.8 & 0.78\\
  3115 & S0   &   9.7 & 4 & 229 & 4 &    --    &    --  &    --   & 34.6 & 20    & 249.4 & 0.61 &  \ 9.0 & 0.54 \\
  3379 & E1   & 10.6 & 1 & 197 & 1 & 4.03 & -0.11 & 0.29 & 40.1 & 10     & 191.7 & 0.15 &  \ 7.5 & 0.75\\
  4278 & E12 & 16.1 & 1 & 228 & 1 & 4.15 & -0.06 & 0.40 & 31.5 & 10     & 155.0 & 0.07 &  \ 7.7 & 0.66\\
  4472 & E2   & 16.7 & 2 & 288 & 1 & 3.87 & -0.22 & 0.30 & 94.9 & 20     & 313.4 & 0.19 & 41.6 & 0.54\\
  4594 & SA   &   9.0 & 3 & 251 & 3 &    --   &    --   &    --   & 70.2 & 22.5* & 297.1 & 0.46 & 18.0 & 0.42\\
  4649 & E2   & 16.5 & 2 & 308 & 1 & 4.23 & -0.12 & 0.36 & 66.4 & 20     & 241.3 & 0.19 &  29.6 & 0.61\\
  4697 & E6   & 11.7 & 1 & 180 & 1 & 3.30 & -0.29 & 0.26 & 62.3 & 10     & 240.2 & 0.37 &  \  8.3 & 0.81\\
  7457 & SA0& 13.2 & 4 &  \ 74& 1 & 2.77 & -0.19 & 0.12 & 36.5 &  \ 5    & 155.1 & 0.45  &  \ 2.0 & 0.90\\
\enddata
\tablenotetext{a}{galaxy classifications from \citet{deVaucouleurs91} } 
\tablenotetext{b}{distances in Mpc derived from surface brightness fluctuation measurements by: (1) \citet{Blakeslee01}; (2) \citet{Blakeslee09}; (3) \citet{Jensen03}; (4) \citet{Tonry01} }
\tablenotetext{c}{velocity dispersion ($\sigma$) from: (1) \citet{Cappellari12}; (2) \citet{Saglia00}; (3) \citet{Jardel11};  (4) \citet{van_den_Bosch16} and \citet{Emsellem99} }
\tablenotetext{d}{\Mgb lick index, metallicity and $\alpha$ abundance from \citet{McDermid15}. }
\tablenotetext{e}{The effective radius ($R_{e}$), derived using the formulation of \citet{Cappellari11}: the average of the B-band $R_{e}$ from \citet{deVaucouleurs91} and that based on 2MASS LGA data, $R_{e,2MASS}=1.7\times median({\rm j\_r\_eff,h\_r\_eff,k\_r\_eff})\sqrt{\rm k\_ba}$. }
\tablenotetext{f}{The radius defining the central region that is excluded from our analysis *For NGC~4594 we remove an elliptical inner region with semi-minor axis = 22.5\arcsec and semi-major axis = 168\arcsec. }
\tablenotetext{g}{Galaxy data from the two micron all sky survey (2MASS) large galaxy atlas (LGA) \citep{Jarrett03}, `total' extrapolated galaxy semi-major axis (${\rm r_{ext}}$), ellipticity ($e=1-b/a$), total $K$-band luminosity within this ellipse ($L_{K}$) assuming $M_{K\odot}=3.33$. }
\tablenotetext{h}{fraction of $L_{K}$ covered by this study, i.e. in the region $r_{in}<r<r_{ext}$ and covered by the \hst and \chandra observations. }
\vspace{-2.0em}
\end{deluxetable*}

Stellar kinematics have also been used to propose a variable IMF in early-type galaxies. Observations of the fundamental plane \citep[e.g.][]{Dressler87, Djorgovski87} have shown that the $M/L$ increases systematically with $\sigma$, suggesting that the dark matter fraction and/or IMF vary systematically with $\sigma$ \citep[e.g.][]{Renzini93, Zepf96, Graves10, Dutton12, Cappellari13}. Through detailed modeling of IFU data, \citet{Cappellari12,Cappellari13} were only able to explain this variation with an IMF model that has a $M/L$ ratio that increases with $\sigma$. These dynamical results (as well as those from gravitational lensing) can be explained by an IMF which becomes increasingly bottom-heavy (where more low mass stars add mass but relatively little light), increasingly top-heavy (where an increased number of stellar remnants add additional mass) or, perhaps, both. 

The stellar remnant fraction can also provide a test for variations in the IMF. This can be probed via low mass X-ray binaries, which consist of accreting neutron stars and black holes \citep{Weidner13b, Peacock14}. These LMXBs provide one of the few direct constraints on the massive end of the IMF in old stellar populations. \citet[][hereafter P14]{Peacock14} used the LMXB populations of seven local early-type galaxies to show that a variable IMF, which varies from Kroupa like at low galaxy mass to steep power-laws at high galaxy mass, is inconsistent with the observed populations. However, these constraints were limited by a lack of data for low mass galaxies. Recently, \citet[][hereafter C17]{Coulter17} published further constraints on IMF variations based on LMXB populations. By considering galaxies with shallow X-ray data, they were able to include more galaxies, at the expense of having few LMXBs per galaxy and only detecting the brightest sources (those with $\lx>10^{38}$~\ergs). They confirmed the finding of P14 and extended this work to consider more complex forms of IMF variations. In this paper, we extend this work to consider additional data and models.

\section{LMXB populations of local early-type galaxies} 
\label{sec:data}

Our approach to constraining variations in the IMF is to utilize the field LMXB populations of galaxies. Our sample of galaxies and their field LMXB populations are taken from P14, \citet{Lehmer14} and \citet{Peacock17b}. We review these data here, but refer the reader to those papers and the citations therein for further details. 

Our sample consists of local ($<20$~Mpc) early-type galaxies with deep \chandra observations ($>100$~ks). From these \chandra data a low contamination sample of field LMXBs is produced by using \hst data to remove globular cluster LMXBs and background AGN. Throughout this paper we consider the number of LMXBs in these galaxies to be the number with $\lx > 2\times10^{37}\ergs$ (a factor of five times deeper than the sample studied by C17). Six of the galaxies in our sample are complete to this limit (NGC~3115, NGC~3379, NGC~4278, NGC~4594, NGC~4697, NGC~7457). We also include NGC~4649, NGC~4472 and NGC~1399 whose field LMXB populations are large enough to reliably extrapolate their size to $2 \times 10^{37} \ergs$, assuming a universal XLF. For all galaxies we calculate the number of field-LMXBs with $\lx > 2\times10^{37}\ergs$ based on fitting to this XLF. The galaxy XLFs and details of this fitting are presented in Appendix \ref{sec:appendix}. 

The number of LMXBs in each galaxy is scaled by the K-band stellar luminosity covered by these \chandra and \hst data based on photometry from the 2MASS LGA \citep{Jarrett03}. We take the outer extent of each galaxy to be the $r_{ext}$ ellipse, as defined by the 2MASS LGA catalog. We exclude the innermost regions of each galaxy, with $r>10\arcsec-20\arcsec$ ($r_{in}$). This is because the X-ray sensitivity is lower in the central regions due to crowding and/or a hot gas component. The reliability of source classification is also lower in the crowded core regions. Table \ref{tab:gal_data} summarizes the properties of the galaxies in our sample.

\begin{figure}
\epsscale{1.1}
\plotone{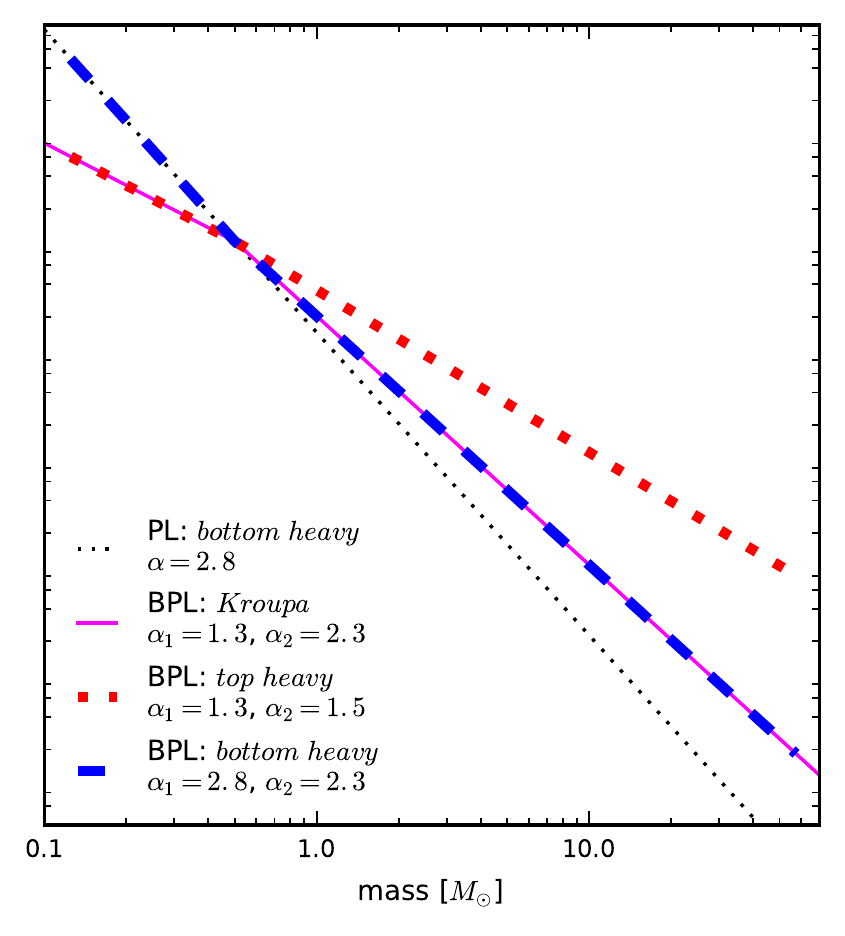}
 \caption{The IMF forms considered in this paper (scaled to have a similar number of stars with $m=0.5\Msun$). The solid-magenta line shows a Kroupa like IMF. This is similar to that observed in the Milky Way and consists of a broken powerlaw (BPL) with $\alpha_{1}=1.3$ and $\alpha_{2}=2.3$. The dashed-blue line shows a bottom heavy BPL model which is similar Kroupa above 0.5\Msun, but has a steeper slope at lower stellar mass, with $\alpha_{1}=2.8$. The dotted-red line shows a top heavy BPL model which is similar Kroupa below 0.5\Msun, but has a shallower slope at higher stellar mass, with $\alpha_{2}=1.5$. The dotted-grey line shows a bottom heavy single powerlaw (PL) IMF with exponent, $\alpha=2.8$. 
 \label{fig:IMFs} }
\end{figure}

\section{The number of field LMXBs} 

In this Section, we compare the specific frequency of field LMXBs in these galaxies, \nx = number LMXBs per $10^{10}$ \LKsun, to predictions from different IMF models. The number of LMXBs is expected to scale directly with the number of neutron stars and black holes in a galaxy. Therefore, \nx probes the ratio of the initial number of stars with masses $>8$~\Msun to the number of stars currently dominating the galaxy's K-band emission. 

Below, we consider the predicted variation of \nx with $\sigma$ for an invariant IMF and IMFs which become either increasingly top or bottom heavy with increasing $\sigma$. For reference, Figure \ref{fig:IMFs} compares the top and bottom heavy IMFs that we consider to a Kroupa IMF. We note that many other forms of variable IMF models could be constructed -- for example, with multiple breaks, breaks at different stellar masses, or varying multiple components of the IMF. We do not have sufficient data to meaningfully constrain all of this parameter space, so we restrict our discussion to only these forms.

\subsection{An invariant IMF} 

Under an invariant IMF, the ratio of high to low mass stars should be similar among the galaxies. This will result in similar fractions of compact objects and, therefore, a similar \nx across these galaxies, which all have old stellar populations. 

\begin{figure}
\epsscale{1.2}
\plotone{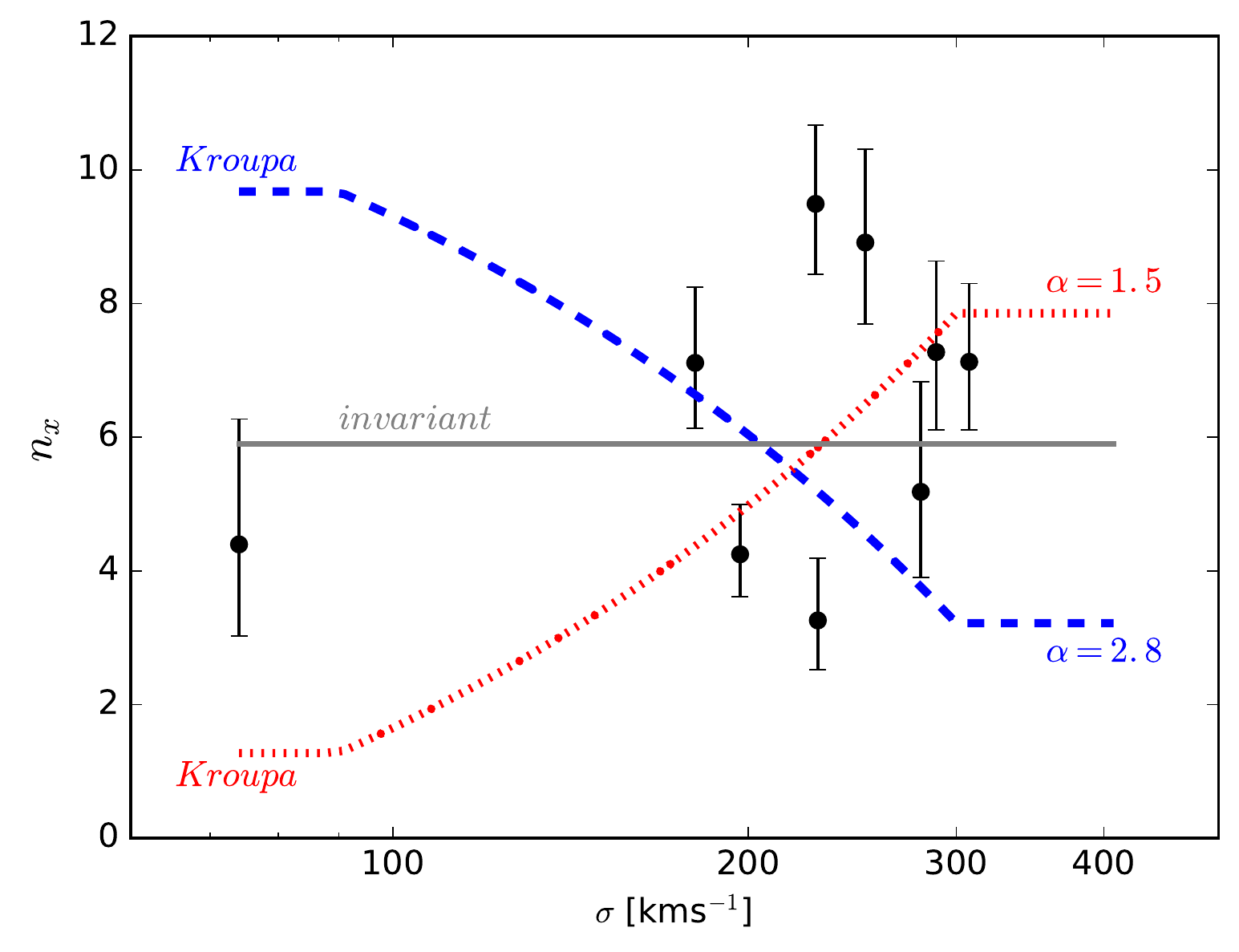}
 \caption{The specific frequency of LMXBs, $n_{x} = \#LMXBs/10^{10}\LKsun$, as a function of velocity dispersion ($\sigma$, black points). The lines compare these data to the predictions presented in P14 for an invariant IMF (solid grey line), an IMF which varies from Kroupa at low $\sigma$ to a single power law with $\alpha=2.8$ at high $\sigma$ (dashed blue line), and an IMF which varies from Kroupa at low $\sigma$ to a single power law with $\alpha=1.5$ at high $\sigma$ (dotted red line). The formation efficiency of LMXBs is poorly constrained theoretically. We therefore scale all models to fit the data (and hence predict different \nx at low $\sigma$, where all models have a similar IMF). 
 \label{fig:P14} }
\end{figure}

Figure \ref{fig:P14} shows \nx as a function of $\sigma$ for the nine galaxies studied (black points). The solid grey line is the prediction for an invariant IMF. The formation efficiency of LMXBs is poorly constrained theoretically. We therefore fit the scaling of this (and subsequent) IMF models to the data. This constant IMF provides a reasonable representation of the data, with no clear trends observed with $\sigma$. However, significant scatter is observed with $\chi^{2}/\nu$ = 3.8 (for $\nu=8$). This scatter can not be explained by an IMF that varies systematically with $\sigma$. However, it suggests that either another factor influences the formation of these binaries or there is an additional source of error, beyond the Poisson noise considered. 

The variation in the LMXB populations of these galaxies is different to the conclusion of P14, where a similar \nx was proposed with NGC~3379 the only significant outlier. C17 also found that \nx was consistent with a constant value for a slightly larger sample of galaxies, although their uncertainties for individual galaxies were larger due to a brighter X-ray detection limit of $\lx>10^{38}$~\ergs. The increased variation compared with P14 is driven by the inclusion of NGC~3115 which has a well constrained and even lower \nx than NGC~3379. As noted above, this can not be explained by a correlation with $\sigma$. Future observations will allow us to measure \nx for larger samples of galaxies and test for correlations with other parameters (see Sections \ref{sec:Z} and \ref{sec:a_dyn}).

\subsection{Variations in the IMF with $\sigma$} 

As discussed in Section \ref{sec:intro}, independent studies based on spectral fitting, stellar dynamics, and gravitational lensing all suggest that the IMF may vary systematically with $\sigma$. Observations of large numbers of early-type galaxies have demonstrated a strong correlation between the mass to light ratio ($M/L$) of a galaxy and its velocity dispersion \citep[$\sigma$; e.g.][and the references therein]{Graves10, Cappellari13}. This trend is larger than expected due to stellar population differences, such as metallicity \citep[][and references therein]{Graves10}. Two explanations are commonly presented to explain this trend. It is thought that higher $\sigma$ galaxies have higher dark matter fractions and/or IMFs with higher $M/L$ ratios. Through detailed modeling of SAURON IFU spectra of 260 early-type galaxies, \citet{Cappellari12, Cappellari13} argued that dark matter fractions cannot fully explain this trend. 

In this Section we assume that the $M/L$ increases due to a systematically varying IMF, as proposed by \citet{Graves10} where $M/L\propto\sigma^{0.65}$. \citet{Cappellari13} also propose that the IMF must vary with $\sigma$ to explain the observed $M/L$ ratio, although they find a smaller exponent. We use the \citet{Graves10} relation because the proposed variation in the IMF is more consistent with the variation proposed to explain the observed spectroscopic abundances. 

Figures \ref{fig:P14} -- \ref{fig:bpl_top}, show \nx as a function of $\sigma$ for our sample of galaxies and compares this to the predictions from different variable IMF models, which we describe below.

\subsubsection{Kroupa $\rightarrow$ bottom heavy power-law IMF}

In Figure \ref{fig:P14} we compare \nx as a function of $\sigma$ to the predictions from the IMF model presented in P14 (dashed-blue line). In this model the IMF becomes more bottom heavy with increasing $\sigma$. This model requires that galaxies with $\sigma<95 \kms$ have a Kroupa like IMF and galaxies with $\sigma>300 \kms$ have a single (bottom heavy) power-law IMF with $\alpha=2.8$. These values are consistent with the IMFs proposed from both spectroscopic studies \citep[e.g.][]{vanDokkum10} and dynamical studies \citep[e.g.][]{Cappellari12}. For these two IMFs we estimate the observed $M/L$ ratio based on the models of \citet[][re-run for this power-law IMF]{Maraston05}. As $\sigma$ increases from 95 to 300~\kms we increase the fraction of stars formed with the $\alpha=2.8$ IMF such that $M/L\propto\sigma^{0.65}$. For further details of this model, please see P14. 

This single power-law bottom heavy IMF model is inconsistent with the data with $\chi^{2}/\nu$ = 7.7 (for $\nu=8$). This is similar to the conclusion of P14, but based on stronger constraints over a broader range of $\sigma$. It is also similar to conclusions drawn from considering the brightest LMXBs in a slightly larger sample of galaxies (C17).

\subsubsection{Kroupa $\rightarrow$ top heavy power-law IMF}
\label{sec:P14t}

Proposed variability in the IMF from stellar absorption lines generally requires an IMF which becomes increasingly bottom heavy with $\sigma$. However, IMF variability inferred from dynamical and gravitational lensing observations can also be explained by an IMF which becomes increasingly top heavy with $\sigma$. For these top-heavy IMFs, the $M/L$ ratio is higher for high mass galaxies due to the increased number of stellar remnants produced. 

The dotted-red line in Figure \ref{fig:P14} shows this increasingly top-heavy IMF model. This is produced following a similar process to that of P14, but where the IMF varies from a Kroupa IMF at $\sigma<95 \kms$ to a single power-law with $\alpha=1.5$ for $\sigma>300 \kms$. 

Interestingly, this top-heavy single power-law model provides a reasonable representation of these data. This model predicts significantly more LMXBs in high mass galaxies. But, with $\chi^{2}/\nu$ = 4.0 (for $\nu=8$), the fit is similar to that of the invariant IMF model. Given the scatter observed around both models, the addition of more galaxies is required to distinguish between the distinctly different predictions of these two IMF models.

\subsubsection{Kroupa $\rightarrow$ bottom heavy broken power-law IMF}
\label{sec:bpl_bottom}

The spectroscopic evidence for IMF variations has generally focused on lines that are strong in low mass stars. Therefore, while single power-law IMFs have sometimes been invoked to explain observations of the most massive galaxies, the observations only require an increased number of low mass stars ($<0.5\Msun$). This means that broken power-law IMFs (or equivalent forms) can explain the observations by only varying the fraction of low mass stars. In this section we consider a Kroupa like IMF which has the form: 

\begin{equation}
 \xi(m) \propto \begin{cases}
   \hfill m^{-\alpha_{1}}, & \text{$0.1M_{\odot}<m<0.5M_{\odot}$}\\
    2^{(\alpha_{2}-\alpha_{1})}\ m^{-\alpha_{2}}, & \text{$0.5\Msun<m<100M_{\odot}$}\\
 \end{cases}
 \label{eq:bpl}
\end{equation}
\\
Here, $\xi(m)$ is the number of stars with mass $m$. For a Kroupa IMF $\alpha_{1}=1.3$ and $\alpha_{2}=2.3$. 

C17 explored the combined constraints of the LMXB populations and spectroscopic observations. They demonstrated that, while the LMXB populations place strong constraints on $\alpha_{2}$, all of the data can be explained if only $\alpha_{1}$ varies significantly. We construct a similar variable bottom heavy IMF model where $\alpha_{2}$ is fixed at 2.3 and $\alpha_{1}$ varies to explain the observed $M/L$ -- $\sigma$ relation. 

We produce stellar populations models using the FSPS models \citep{Conroy09, Conroy10}, implemented under pythonFSPS \citep{Foreman-mackey14}. We run these models with the MILES spectral libraries and the PADOVA isochrones for a simple stellar population with solar metallicity and an age of 10~Gyr. We produce a group of models for IMFs with $\alpha_{2}$=2.3 and $\alpha_{1}=1.3, 1.4, ..., 2.8$. Each model generates a $M/L$ ratio for the stellar population which includes the mass of stellar remnants. 

\begin{figure}
\epsscale{1.2}
\plotone{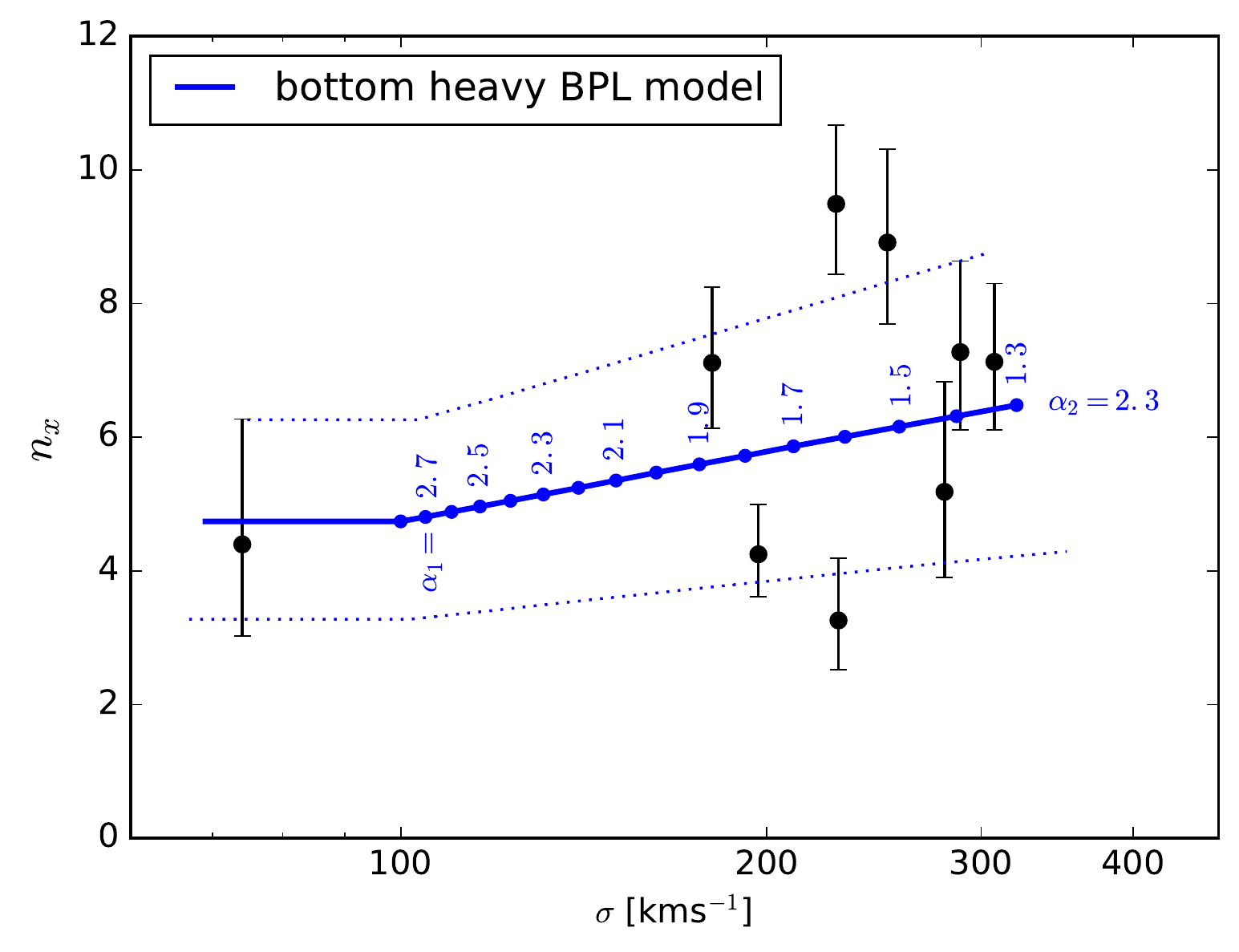}
 \caption{Data as in Figure \ref{fig:P14}. The solid-blue line shows an increasingly bottom heavy IMF model in which the number of low mass stars ($<0.5\Msun$) increases systematically with $\sigma$, with $\alpha_{1}$ increasing from 1.3 to 2.8 ($\alpha_{2}$ remains constant at 2.3). The dotted-blue lines are for $\alpha_{2}=-2.16$ and -2.48, which are the conservative constraints these data place on variation in the high mass slope. 
 \label{fig:bpl_bottom} }
\end{figure}

For each model, we calculate the $\sigma$ associated with its $M/L$ by assuming the observed relation $M/L=c_{kro}\ \sigma^{0.65}$. Here, the constant $c_{kro}$ is calculated by assuming that galaxies with $\sigma=100\kms$ have a Kroupa like IMF, then $c_{kro}=(M/L)_{kro}/(100^{0.65})$. We then calculate the fraction of compact objects (black holes and neutron stars, $f_{CO}$) in the stellar population of each IMF  via: 

\begin{equation}
f_{CO} = \frac {\int^{100}_{8} m\ \xi(m)\ dm} {\int^{100}_{0.1} m\ \xi(m)\ dm}
 \label{eq:f_co}
\end{equation}
\\
Here $\xi(m)$ is taken from Equation \ref{eq:bpl} and the extra mass term is to calculate $f_{CO}$ as a function of total stellar mass. Using Equation \ref{eq:f_co}, we can calculate \nx for each IMF via: 

\begin{equation}
\nx = c_{x} f_{CO} \frac{M}{L_{K}}
 \label{eq:nx}
\end{equation}
\\
Where $M/L_{K}$ is calculated for each IMF and is required to compare $f_{CO}$ with \nx, since this is scaled by the total K-band light, rather than the total mass. The constant of proportionality ($c_{x}$) relates to the formation efficiency of LMXBs and should be similar for all of the IMFs. As with all models considered, we leave this constant as a free parameter which we fit to the data. The resulting grid of models predict how \nx should vary with $\sigma$ for increasingly bottom heavy IMFs. 

\begin{figure}
\epsscale{1.2}
\plotone{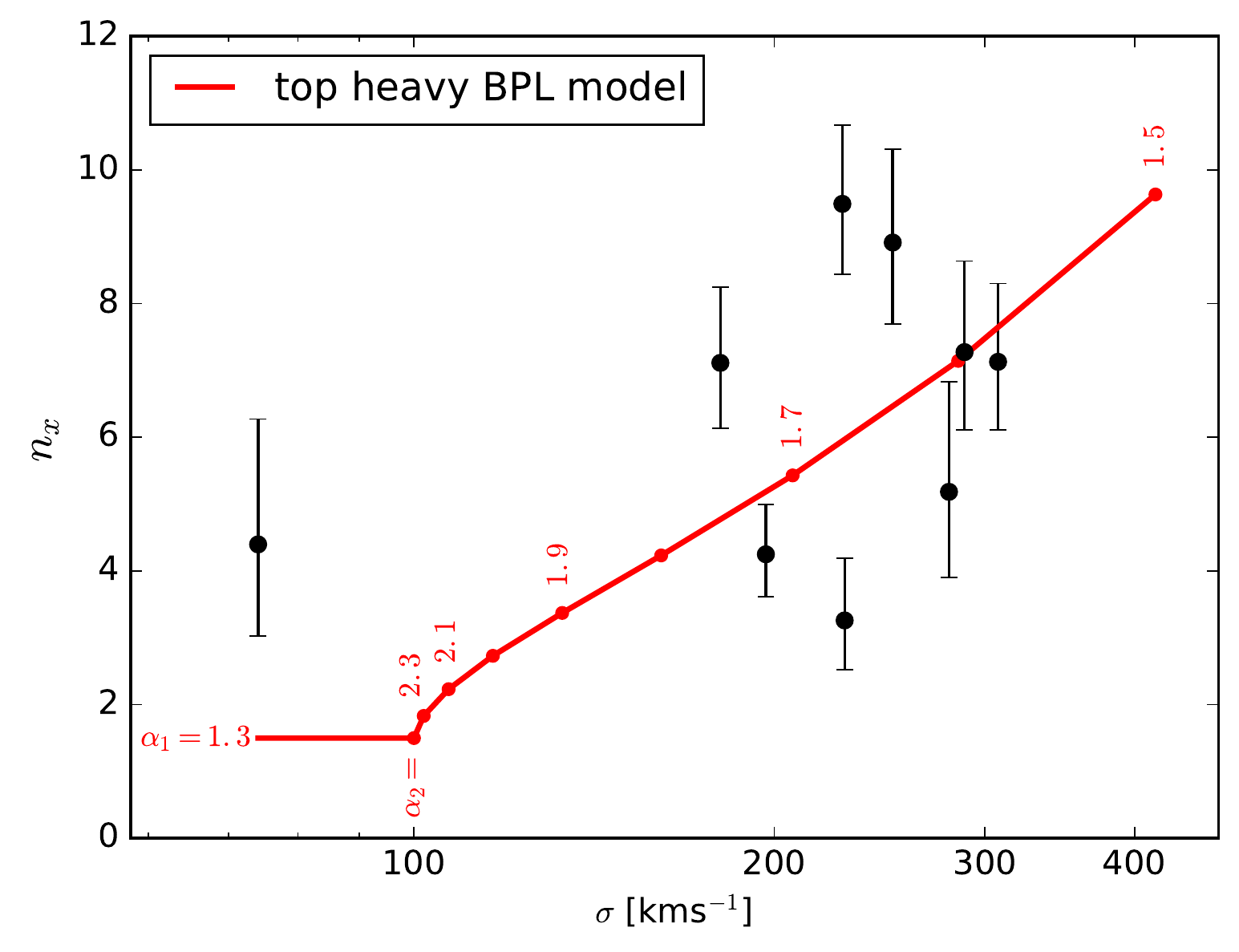}
 \caption{Same as Figure \ref{fig:bpl_bottom}, but where the model is based on an increasingly top heavy IMF. In this model, the number of low mass stars ($m<0.5\Msun$) is constant (with $\alpha_{1}=1.3$) and the number of high mass stars increases with $\sigma$ (with $\alpha_{2}$ varying from 2.3 to 1.5). 
 \label{fig:bpl_top} }
\end{figure}

The solid-blue line in Figure \ref{fig:bpl_bottom} shows this bottom heavy model. The high mass end of the IMF is fixed, with $\alpha_{2}=2.3$ and the low mass end becomes increasingly bottom heavy with $\sigma$. The blue dots along this line show $\alpha_{1}=1.3, 1.4, ..., 2.8$ from left to right, respectively. We note that, while these models were constructed to explain the $M/L$ variation, $\alpha_{1}=2.8$ for the most massive galaxies is also consistent with spectroscopic observations of these galaxies. 

This model provides a reasonable representation of the data and, with $\chi^{2}/\nu = 3.6$ ($\nu=8$), the fit is similar to the invariant IMF model. In this model, the observed $M/L$ ratio trend is accounted for by varying the low mass end of the IMF (which also explains the spectroscopic observations). Given this, we can constrain the permissible variation in the high stellar mass end of the IMF. We fix the scaling of \nx to that found for the $\alpha_{2}=2.3$ IMF and construct a grid of variable $\alpha_{1}$ models for a range of fixed $\alpha_{2}$. We define the range of permissible $\alpha_{2}$ models to be those which have $\chi^{2}<=2\chi^{2}(\alpha_{2}=2.3)$. This constrains $\alpha_{2}=2.30^{+0.18}_{-0.14}$, illustrated by the dotted-blue lines in Figure \ref{fig:bpl_bottom}. 

We note that by making the scaling of \nx a free parameter, we constrain the relative number of LMXBs for a given IMF (and hence the relative number of compact objects). We choose to fix $\alpha_{2}$ at 2.3, which is consistent with the stellar populations observed in the Milky Way. However, fixing $\alpha_{2}$ at other values mainly changes the scaling of the \nx -- $\sigma$ relationship and produces only small changes in the trend observed in Figure \ref{fig:bpl_bottom}. Therefore, since the scaling of \nx is a free parameter, similar quality fits can be obtained for a wide range of $\alpha_{2}$. The LMXB population provides strong constraints on the allowable variation in $\alpha_{2}$, rather than its actual value.

\subsubsection{Kroupa $\rightarrow$ top heavy broken power-law IMF}
\label{sec:bpl_top}

We also consider a variable broken power-law IMF model that is similar to that presented in Section \ref{sec:bpl_bottom} (above), but  where the low mass end of the IMF is fixed (at $\alpha_{1}=1.3$) and the high mass end ($\alpha_{2}$) varies with sigma to explain the observed $M/L - \sigma$ relation. This model is shown as the red line in Figure \ref{fig:bpl_top}. The dots along this line show $\alpha_{2}=2.3,2.2,...,1.5$ from left to right, respectively. 

We note that because top-heavy IMF models require significant variations in the number of compact objects produced, they generally produce significant variation in \nx. This broken power-law top-heavy IMF model predicts a comparable trend to that of the single power-law top-heavy ($\alpha=1.5$) IMF model (see Figure \ref{fig:P14}) and provides a similar quality fit to the data with $\chi^{2}/\nu = 3.8$ ($\nu = 8$). We draw similar conclusions to those of Section \ref{sec:P14t}; that we are currently unable to significantly distinguish between an invariant IMF and this variable top-heavy IMF model. However, the predicted \nx variations are significant and should be distinguishable with the addition of data for more galaxies.

\subsection{Variations in the IMF with metallicity/ stellar abundance} 
\label{sec:Z}

\begin{figure}
\epsscale{1.2}
\plotone{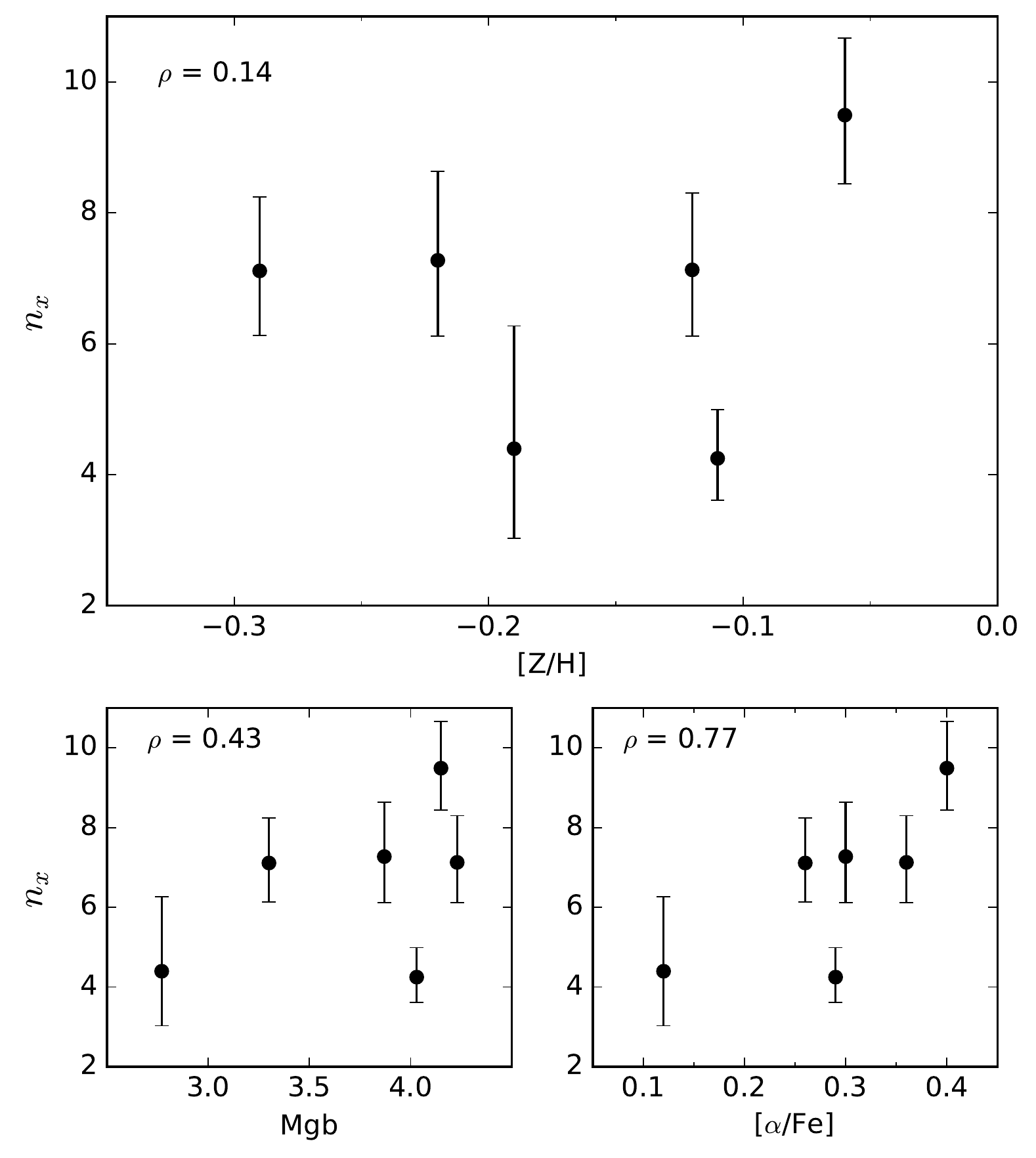}
 \caption{Number of LMXBs with $\lx>2\times10^{37}$ \ergs (\nx) as a function of metallicity ($[Z/H]$), \Mgb and $\alpha$-abundance ($[\alpha/Fe]$). The stellar populations data are calculated from \citet[][measured for light within 1$R_{e}$]{McDermid15}. In the top left of each panel we show the spearman rank order correlation coefficient $\rho$. \nx appears to correlate with $[\alpha/Fe]$, but for this relatively small sample this is not significant.  \label{fig:N_galParam} }
\end{figure}

It has also been proposed that the IMF may vary with stellar abundance. While the presence and magnitude of such correlations vary in different studies, the IMF may vary with metallicity ($[Z/H]$; \citealt{Martin_Navarro15b, vanDokkum16}; although \citealt{McDermid14} found no significant correlation), $[Mg/Fe]$ (\citealt{Conroy12b}; although \citealt{LaBarbera15} found no significant correlation) and $\alpha$-element abundance ($[\alpha/Fe]$; \citealt{Conroy12b, McDermid14}). Indeed, \citet{vanDokkum16} recently concluded that, once radial variations are accounted for (see below), the IMF correlates most strongly with metallicity, rather than $\sigma$ or $[Mg/Fe]$. 

Homogeneous stellar populations data are available for six of the nine galaxies in our sample from the study of  \citet{McDermid15}. They derive mass-weighted parameters from fitting the integrated light within 1$R_{e}$. In Figure \ref{fig:N_galParam}, we plot \nx as a function of $[Z/H]$, \Mgb and $[\alpha/Fe]$ \citep[taken from][]{McDermid15}. Our sample of galaxies are selected to span only a small range of $[Z/H]$ and no significant trend with \nx is observed over this range ($-0.3<[Z/H]<0.0$). Spearman rank tests (see Figure \ref{fig:N_galParam} for values of $\rho$) suggest positive correlations between \nx and both \Mgb and $[\alpha/Fe]$ and it is perhaps interesting that NGC 4278 has a significantly higher \nx 
and the highest $[\alpha/Fe]$. However, neither correlation is significant due to the small sample size. Larger samples will be required to provide a significant test for correlations between \nx and the abundance of light 
elements. We also note that the radial range covered by our LMXB analysis is not identical to that used in the stellar metallicity analysis. Specifically, the metallicity study goes from the center to $1R_{e}$, while our LMXB analysis excludes $r < 0.2 R_{e}$ and extends to slightly larger radii. We therefore caution that significant differences among the metallicity gradients in these galaxies may effect our conclusions regarding metallicity effects.

\subsection{Direct comparison to the IMF mismatch parameter, $\alpha_{dyn}$} 
\label{sec:a_dyn}

 \begin{figure}
\epsscale{1.2}
\plotone{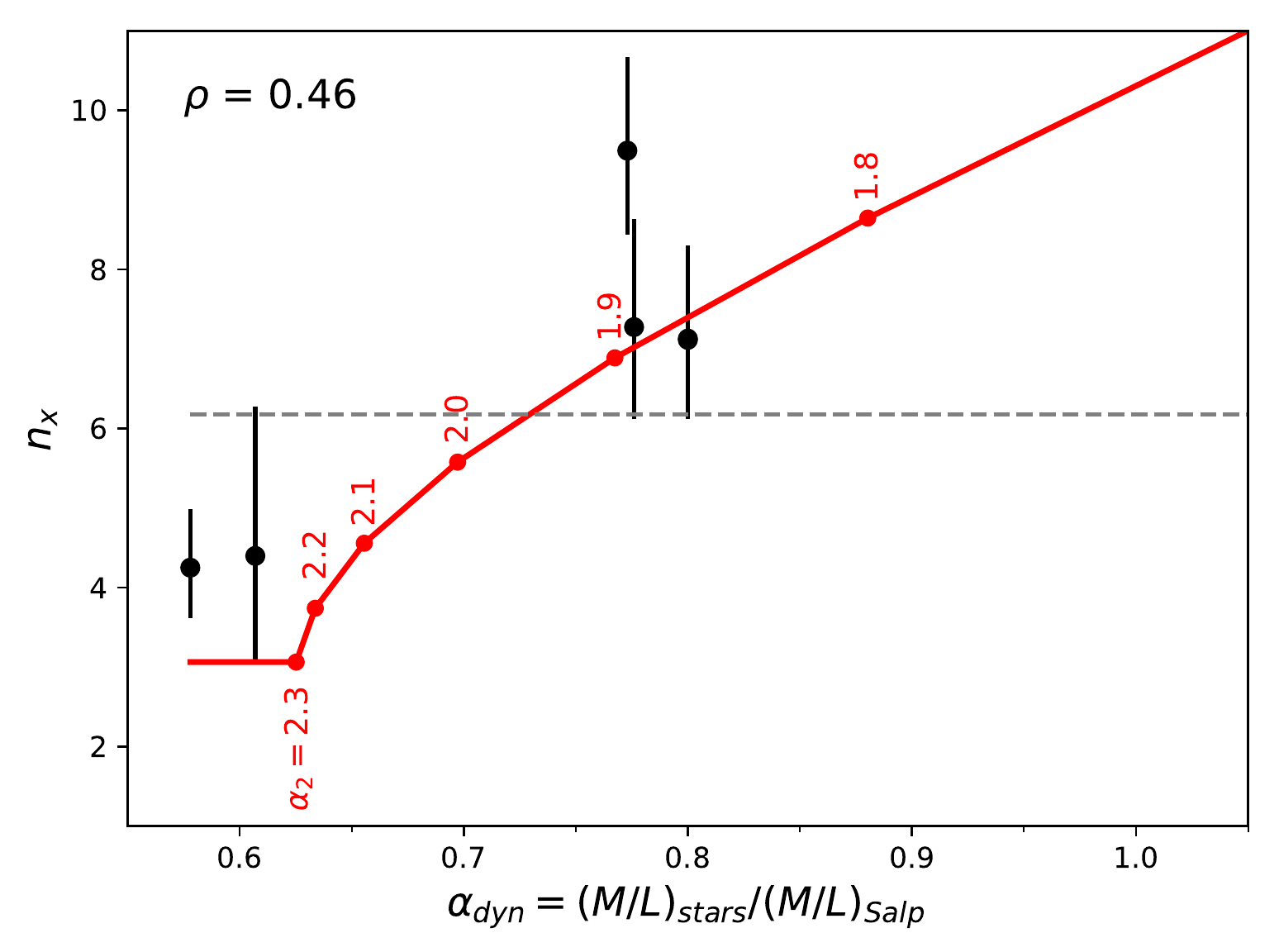}
 \caption{\nx as a function of the ``IMF mismatch" parameter ($\alpha _{dyn}$). $\alpha _{dyn}$ is taken from \citet{Cappellari13} and is based on comparing the $M/L$ ratios calculated via dynamical modeling to those calculated via stellar population synthesis modeling. The red line shows the variation of \nx with $\alpha _{dyn}$ for a variable IMF with $\alpha _{1}$ fixed at 1.3 and $\alpha _{2}$ varying from 2.3 to 1.8 (red points, from left to right).  \\ \label{fig:N_aDyn} }
\end{figure}

For six of the nine galaxies in our sample, it is possible to compare their \nx directly to their ``IMF mismatch" factor ($\alpha_{dyn}$) derived from dynamical modeling of data from the Atlas3D survey \citep{Cappellari12, Cappellari13}. This parameter compares the $M/L$ ratio calculated from dynamical modeling to that derived from stellar population modeling assuming a Salpeter IMF, i.e. $\alpha_{dyn}=(M/L)_{stars}/(M/L)_{Salp}$. 

Figure \ref{fig:N_aDyn}, compares \nx with $\alpha_{dyn}$. Two of the six galaxies in this reduced sample (NGC~4472 and NGC~4649) have high $\sigma$ and are therefore expected to have extreme IMFs. However, their $\alpha_{dyn}$ measurements are quite modest. Our sample varies from $\alpha_{dyn}\sim0.6$ (consistent with Kroupa-like IMFs) to $\alpha_{dyn}\sim0.8$, suggesting an IMF with a higher $M/L$ ratio, but less than that of a Salpeter IMF (which is defined to have $\alpha_{dyn}=1.0$). 

Interestingly, the two lowest \nx galaxies, NGC~3379 and NGC~7457, do have lower $\alpha_{dyn}$ measurements than the four other galaxies. A Spearman rank test suggests that this is not significant. However, our sample is quite small and quite bimodal in $\alpha_{dyn}$, limiting the conclusions from such a test. 

If higher $\alpha_{dyn}$ is the result of a more bottom heavy broken power-law IMF, such as that discussed in Section \ref{sec:bpl_bottom}, we expect little correlation between $\alpha_{dyn}$ and \nx. However, a positive correlation between $\alpha_{dyn}$ and \nx would be expected if higher values of $\alpha_{dyn}$ are the result of more top-heavy IMFs. This would result in more stellar remnants, producing the higher $M/L$ ratios and relatively more LMXBs. The solid-red line in this figure shows the effect of an increasingly top heavy IMF on \nx. Here, we plot $\alpha_{dyn}$ using the same FSPS models discussed in Sections \ref{sec:bpl_bottom} and \ref{sec:bpl_top}. We vary the IMF by fixing $\alpha_{1}$ at 1.3 and increasing $\alpha_{2}$ from a Kroupa-like IMF (with $\alpha_{2}=2.3$) to a top heavy IMF(with $\alpha_{2}=1.8$). The red points along this line indicate $\alpha_{2}=2.3, 2.2, ..., 1.8$, from left to right. For each IMF we predict \nx based on the relative fraction of stellar remnants produced (see Equations  \ref{eq:f_co} and \ref{eq:nx}) and scale the resulting function to fit the data. Such a model provides a good fit to the observed \nx with $\chi^{2}/\nu = 1.53$ ($\nu=5$). 

Directly comparing $\alpha_{dyn}$ to \nx provides an important constraint on potential IMF variations. The agreement between these LMXB populations and the top-heavy IMFs that could be inferred from dynamical modeling is interesting. However, our sample is currently quite small and the addition of new data will provide sharper tests in the future. In particular, \citet{Cappellari13} found massive galaxies with $\alpha_{dyn}>1.0$, suggesting that they may have the most extreme IMFs.

\subsection{Radial variations in the IMF} 

Recent work has proposed that there may be significant radial variations in the IMFs of massive galaxies, with the inner regions ($\lesssim 0.25 R_{e}$) having extremely bottom heavy IMFs and outer regions being more similar to a Milky-Way like IMF \citep[e.g.][]{Martin_Navarro15a, McConnell16, vanDokkum16, LaBarbera16, LaBarbera17}. 

Unfortunately, it is hard to study such variations with the LMXB data utilized in this paper. This is because we have to exclude the innermost regions of the galaxies due to source confusion. This limits our study to radii $>(0.2-0.3)R_{e}$.  Additionally, radially binning the data will results in smaller numbers of LMXBs in each bin and hence weaker constraints on variations. 

Previously, C17 considered \nx for a similar sample of galaxies inside and outside of 1~$R_{e}$. They identified no significant evidence for variation over these ranges. However, we note again that this study excluded the innermost regions of the galaxies, so would not be sensitive to IMF variations if they are limited these very central regions. 

We are not able to improve on the radial considerations of C17 in this work, but, for comparison with other work, we note in Table \ref{tab:gal_data} the regions covered for each galaxy. For most galaxies, we cover $(1/4) R_{e} \lesssim r \lesssim 4 R_{e}$ and it is for these regions that our constraints on the IMF are valid. We note that this radial range includes the majority of the galaxy mass (covering $75-90\%$ of the $K-$band light). \\

\subsection{Time variable IMF models}

Some studies have also invoked time variable IMFs to explain observations of both local and high redshift galaxies \citep[e.g.][]{Vazdekis96, Vazdekis97, Dave08, Weidner13a}. In these models, the IMF becomes increasingly bottom heavy with time. \citet{Weidner13b} discussed a bimodal IMF which is top heavy at early epochs before transitioning to a bottom heavy IMF. Under certain formulations of this model, they are able to construct present day IMFs which are bottom heavy but have similar remnant fractions and hence similar fractions of LMXBs. While we cannot distinguish between such an IMF and a time independent IMF, that is constructed to have similar numbers of massive stars, our results are in agreement with those of \citet{Weidner13b}.

\section{Conclusions}
\label{sec:conclusions}

In this paper, we expand on the work of P14 and C17 to help constrain potential variations in the IMF of early-type galaxies based on their LMXB populations. To better constrain proposed IMF models which vary systematically with $\sigma$, we include new data for the relatively low mass galaxy NGC~7457. 

We consider the LMXB populations over the radial range of $(1/4) R_{e} \lesssim r \lesssim 4 R_{e}$ and confirm the conclusions of P14 and C17; that there is no evidence for systematic variations in the high mass end of the IMF. However, we note that significant scatter is present in the number of LMXBs and that even the invariant IMF model is formally inconsistent with the data in the absence of significant intrinsic scatter in the relation. This is different from the conclusion of P14 due primarily to the addition of NGC~3115 to the galaxy sample (which has a well constrained but relatively small LMXB population). 

We consider the bottom heavy IMF model presented in P14, where the most massive galaxies have a single steep power-law IMF with $\alpha=2.8$. Using the improved constraints presented in this paper, we confirm that the variation in the number of LMXBs produced by this steep power-law IMF is inconsistent with the observed populations. We expand on this work to show that broken power-law IMFs in which only the number of low mass stars vary with galaxy mass are consistent with the data. This is similar to the conclusion of C17 and can explain both the LMXB and the spectroscopic observations. 

We also consider variable IMF models in which the IMF becomes increasingly top-heavy with $\sigma$. We show that variable IMFs constructed from both single power-laws and broken power-laws produce a similar variation in the number of LMXBs. Due to the observed scatter in \nx, we are currently unable to distinguish between these top-heavy IMFs and an invariant IMF. However, these top-heavy IMFs predict significantly more LMXBs in massive galaxies and additional data should enable us to distinguish between an invariant IMF and this increasingly top-heavy IMF. 

Recent studies have suggested correlations between the IMF and stellar abundance/ metallicity. We identify no significant trends between \nx and stellar populations parameters, but note that our sample of galaxies with such data is small and spans a limited range of metallicities. In the future, larger samples may allow us to test for such correlations. 

Six of the galaxies in our sample have estimates of their IMFs from dynamically derived $M/L$ ratios, $\alpha_{dyn}$. Assuming that galaxies with higher $M/L$ ratios have top-heavier IMFs, we show that the variation of \nx with $\alpha_{dyn}$ is consistent with these data. However, our sample of galaxies with both \nx and $\alpha_{dyn}$ measurements is small and spans only a limited range of $0.6<\alpha_{dyn}<0.8$. Future observations of galaxies with higher $\alpha_{dyn}$ will enable us to test this interesting correlation further.

\acknowledgements

We thank the anonymous referee for their comments which were beneficial to the final version of this paper. MBP and SEZ acknowledge support from NASA through the ADAP grant NNX11AG12G. BDL acknowledges support from NASA ADAP NNX16AG06G. 

The scientific results reported in this article are based on observations made by the Chandra X-ray Observatory. Support for this work was provided by NASA through Chandra Award numbers GO5-16084A (MBP) and GO5-16084B (AK) issued by the Chandra X-ray Observatory Center, which is operated by SAO for and on behalf of the NASA under contract NAS8-03060. 

Based on observations made with the NASA/ESA Hubble Space Telescope, obtained at the Space Telescope Science Institute, which is operated by the Association of Universities for Research in Astronomy, Inc., under NASA contract NAS 5-26555. These observations are associated with program HST-GO-13942.001-A. Support for program number HST-GO-13942.001-A was provided by NASA through a grant from the Space Telescope Science Institute, which is operated by the Association of Universities for Research in Astronomy, Incorporated, under NASA contract NAS5-26555.

This research has made use of NASA's Astrophysics Data System (ADS) and the NASA/IPAC Extragalactic Database (NED). 

\facilities{CXO, HST, 2MASS}
\software{SciPy, NumPy, matplotlib, pythonFSPS, TOPCAT, ds9}

\bibliographystyle{apj}
\bibliography{bibliography_etal}

\appendix{}

\section{Field LMXB XLF} 
\label{sec:appendix}

\begin{figure*}
 \epsscale{1.0}
\plotone{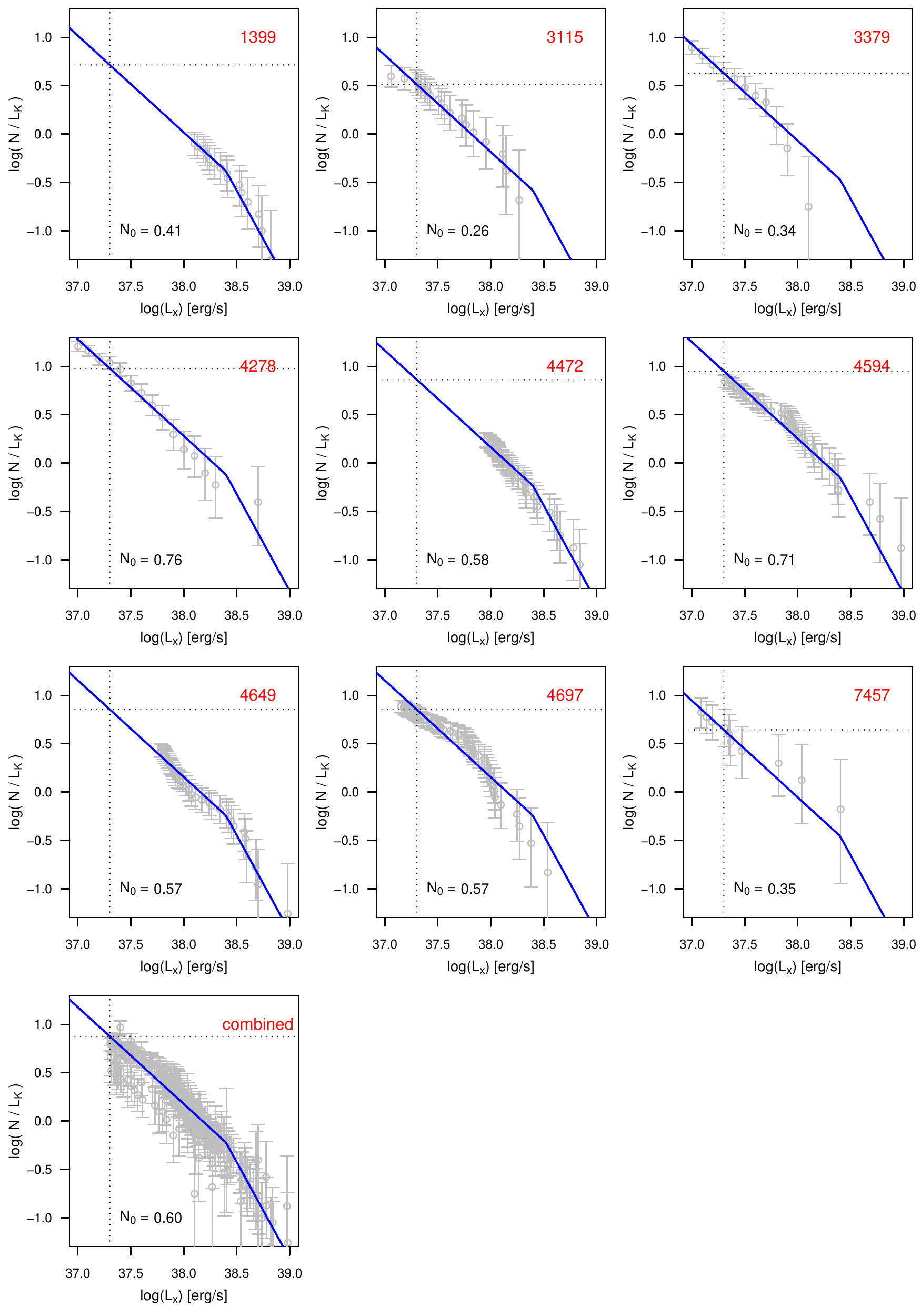}
 \caption{X-ray luminosity functions (XLFs) of field LMXBs in our sample of early-type galaxies. All XLFs are scaled by the K-band stellar light covered. The blue line shows a broken power law that is scaled to fit these data (see text for details). The best fit scaling for each XLF ($N_{0}$) is shown in the bottom left of each panel. The dotted lines show the inferred number of LMXBs with $\lx > 2 \times 10^{37}$~\ergs.  \label{fig:xlfs} }
\end{figure*}

Figure \ref{fig:xlfs} shows the XLFs of each galaxy's field LMXB population. These are based on the field LMXB populations presented in P14 (and references therein), \citet{Lehmer14} and \citet{Peacock17b} and have been scaled by the stellar light covered (see Table \ref{tab:gal_data}). Throughout this paper, we quote the number of LMXBs with $\lx > 2 \times 10^{37}$~\ergs. This luminosity is highlighted with the dotted line. Six of the galaxies are complete to this limit. For the other three galaxies -- NGC~1399, NGC~4472, and NGC~4649 -- we extrapolate the number of LMXBs to fainter magnitudes based on fitting a broken power-law of the form: 

\begin{equation}  \label{equ:bpl}
 \frac{N(>\lx)}{10^{10}\LKsun} = N_{0} \times \begin{cases}
    (\lx/2.5)^{-2.0}, & \text{if $\lx>2.5$}\\
    (\lx/2.5)^{-1.0}, & \text{otherwise}
 \end{cases}
\end{equation} 
\\
Where \lx is in units of $10^{38}$~\ergs and $N_{0}$ is the constant of proportionality. It can be seen that this function provides a reasonable representation of the combined XLFs of all of the galaxies (final panel of Figure \ref{fig:xlfs}). This function is also consistent with previous work \citep[e.g.][]{Kim09, Kim10, Zhang12, Lehmer14, Peacock16}. We fix the exponents at these values and fit the scaling ($N_{0}$) to each galaxy's XLF. For the combined XLF, we find that $N_{0}=0.60$. We note that this updated form of the XLF is slightly flatter than that used in P14, resulting in slightly different extrapolated numbers. Since this form of the XLF better represents those observed, we use these updated values in this paper. 

\label{lastpage}

\end{document}